\documentclass[%
 reprint,
 amsmath,amssymb,
 aps,
]{revtex4-2}

\usepackage{graphicx}
\usepackage{dcolumn}
\usepackage{bm}
\usepackage{hyperref}
\usepackage{braket}
\usepackage{xcolor}
\usepackage{comment}
\usepackage{adjustbox}

\begin{document}

\title{Comparison of non-decoy single-photon source and decoy weak coherent pulse in quantum key distribution}%

\author{Roberto G. Pousa}
 \email{roberto.gonzalez-pousa@strath.ac.uk}
\author{Daniel K. L. Oi}
\author{John Jeffers}
 \affiliation{%
 SUPA Department of Physics, University of Strathclyde, \\
 Glasgow, G4 0NG, United Kingdom
}%

\begin{abstract}
Advancements in practical single-photon sources (SPS) exhibiting high brightness and low $g^{(2)}(0)$ have garnered significant interest for their application in quantum key distribution (QKD). To assess their QKD performance, it is essential to compare them with the widely employed weak coherent pulses (WCPs) in the decoy state method. In this work, we analyze the non-decoy efficient BB84 protocol for an SPS, partially characterising its photon statistics by its $g^{(2)}(0)$ and mean photon number. We compare it to the 2-decoy efficient BB84 with WCPs within the finite-key analysis framework while optimizing the parameters of both protocols. Our findings indicate that the non-decoy SPS with a mean photon number of $\braket{n} = 0.5$ and $g^{(2)}(0) = 3.6\%$ can enhance the secure key generation over the 2-decoy WCP for block sizes under $4.66 \cdot 10^9$ sent signals ($29$ seconds of acquisition time) at a channel loss of $10$ dB ($52.5$ km of optical fibre). Additionally, we demonstrate an increase in the maximum tolerable channel loss for SPSs with mean photon number $\braket{n} \geq 0.0142$ at block sizes below $10^8$ sent signals ($0.62$ seconds of acquisition time). These results suggest that SPSs hold potential for key rate enhancement in short-range QKD networks, though further research is required to evaluate their key generation capabilities when integrated into the decoy method.
\end{abstract}


\maketitle

\textit{Introduction.} Public key cryptosystems have long proposed secure communication schemes between parties using two keys, in which the sender (Alice) uses one key for encryption, randomly applying an operation whose mutually inverse is used for decryption. Although Alice publicly announces the encryption method, the decryption instructions must remain confidential to the intended receiver (Bob). Allowing anyone to encrypt a message, the parties can swap roles and converse secretly. In a classical public key system, it is not possible to guarantee the private key cannot be derived from the encrypted key by computational methods ~\cite{hellman2002overview, grollmann1988complexity}. Seeking a security proof method, quantum mechanics was introduced, leading to the development of the quantum key distribution (QKD) field after a lot of refinement~\cite{bennett1992quantum}.

QKD protocols utilize a quantum channel to transmit signals with complementary physical properties, e.g. in the BB84 protocol, the exploited quantum property to detect an eavesdropper (Eve) is the polarization of light~\cite{bennett2014quantum}. When Eve attempts to monitor the channel measuring the polarization, she inevitably disturbs a certain number of signals due to the Heisenberg uncertainty principle. This causes errors, allowing Alice and Bob to detect Eve's presence and estimate the information leaked to her. By setting an error threshold, Alice and Bob have a criterion to abort the protocol if the error rate exceeds a certain level.

An ideal QKD system would employ a quantum source that consistently sends single-photons. Although ongoing research aims to approach this ideal behaviour ~\cite{wang2019towards}, it is strictly unachievable in practice. Instead, in real QKD systems, this inherent nonideality of practical sources is exploited by the photon number splitting (PNS) attack ~\cite{brassard2000limitations, lutkenhaus2000security}. In a BB84 protocol employing exclusively one source intensity, known as non-decoy, Eve, who fully controls the channel, can perform PNS attacks to effectively intercept the key whilst remaining undetectable by exploiting the multiphotons. By employing quantum nondemolition measurements~\cite{braginsky1980quantum}, Eve identifies the number of photons contained in each pulse. He then blocks the single-photon pulses and splits the multiphoton pulses into single-photons, which are forwarded to Bob, while the remaining photons are stored and measured after basis reconciliation. Therefore, to ensure the security of the non-decoy protocol the click probability of Bob's detectors has to be greater than the multiphoton emission of Alice, $p_{\mathrm{click}} > p_{\mathrm{mp}}$. Otherwise, assuming all multiphoton emissions cause a detection event on Bob's apparatus through Eve's lossless channel, the expected detection rate by Bob can be reproduced by Eve exclusively using the sent multiphotons in a high-loss scenario. Compromising the secrecy of the entire shared key bit string. Consequently, only non-multiphoton detections can be considered secure events.

Weak coherent pulse (WCP) sources, which are attenuated lasers approaching the single-photon regime, have been the most commonly employed sources in prepare-and-measure QKD protocols due to their feasibility~\cite{stucki2005fast}. However, even strongly attenuated WCPs have a sufficiently high multiphoton emission to be vulnerable to PNS attacks in lossy channels, failing to fulfil $p_{\mathrm{click}} > p_{\mathrm{mp}}$. To mitigate such attacks and ensure the inequality is met, the decoy method was proposed~\cite{hwang2003quantum}, which underwent significant refinement~\cite{lo2005decoy,rosenberg2007long,schmitt2007experimental,rusca2018finite}. This technique involves using multiple intensity levels for WCPs with identical characteristics to bound the multiphoton events, i.e. detections caused by multiphotons. While ensuring both parties agree on their single-photon and vacuum events estimate, which form the secure key. This accurate estimation occurs since the used WCPs produce equal counting rates for a sent $k$-photon state, commonly known as yields, despite their distinct photon emission probabilities. In fact, deviations in the yields between different WCP intensities indicate the presence of an eavesdropper, who lacks knowledge regarding the transmitted distribution.

Alternative quantum sources, such as single-photon sources (SPS) based on defects in 2D materials or quantum dots, exhibit significantly lower multiphoton emission than attenuated WCPs~\cite{vogl2017room,morrison2021bright,thomas2021bright}. Therefore, considering no decoy states, these low-$g^{(2)}(0)$ SPSs satisfy the $p_{\mathrm{click}} > p_{\mathrm{mp}}$ condition at higher channel losses than a WCP, emerging as promising candidates for non-decoy protocols. However, to evaluate the QKD performance of practical SPSs, it is necessary to compare it with a state-of-the-art decoy WCP protocol. Thus, this work analyses the key generation of an efficient BB84 protocol for non-decoy SPS, expanding the analysis of~\cite{morrison2023single} to other SPS characteristics plus providing further theoretical aspects, and for 2-decoy WCP based on~\cite{lim2014concise}. Note that efficient BB84 uses one basis (X basis) for key generation and the other (Z basis) for parameter estimation, doubling the efficiency of standard BB84~\cite{lo2005efficient}.

In the decoy method, the single-photon and vacuum events are lower-bounded by analysing the statistics from the implemented multiple source intensities. In contrast, the non-decoy protocol estimates secure non-multiphoton events, lumping together the single-photon and vacuum events, excluding the insecure multiphoton events from the total sifted events, i.e. instances shared by both parties when they chose the same basis. To avoid overestimating the non-multiphoton events, it is essential to upper bound the sent multiphotons and the multiphoton events received by Bob. The subsequent finite-key analysis will address the latter, while Alice's source characterisation will handle the former.

\textit{Source characterisation and multiphoton probability.} Our goal is not to fully characterise the SPS~\cite{alleaume2004photon}, rather, we seek an estimate of the source statistics that upper bounds the multiphoton emissions. We denote the true photon emission probabilities of the SPS in the Fock basis as $\{P_k^{(SPS)}\}_{k \in \mathbb{N}}$ for the infinite Hilbert space, where the multiphoton emission probability is $P_{\mathrm{mp}}^{(SPS)} = \sum_{k \geq 2}  P_k^{(SPS)}$. We assume these true probabilities are not directly accessible. Instead, we estimate the photon number distribution of the SPS using the mean photon number $\braket{n}$ and the time-zero second-order correlation function $g^{(2)}(0)$. Consequently, the multiphoton probability can be upper-bounded as $P_{\mathrm{mp}}^{(SPS)} \leq \overline{p}_{\mathrm{mp}} =  g^{(2)} \braket{n}^2 / 2$~\cite{waks2002security}. Here, uppercase $P$ represents the true emission probabilities, while lowercase $p$ denotes bounded estimates.

To simulate the counts of Bob's apparatus as in a real experiment, we select the set of photon states emitted by Alice's source which reaches exactly $\overline{p}_{\mathrm{mp}}$, ensuring no other combination of states exceeds this upper bound. We study two types of distributions. First, we examine a pathological distribution where all the emission probabilities are null except three: vacuum, single-photon and $K$-photon probabilities $\{p_0, p_1, p_K\}$, where $K \geq 3$ is a fixed value. In this case, all multiphotons are $K$-photon states, and any possible distribution yields a lower upper bound than $\overline{p}_{\mathrm{mp}}$. As expected, when $K$ tends to infinity, the multiphoton probability approaches $\overline{p}_{\mathrm{mp}}$, thus saturating the initial upper bound on the multiphoton probability.

We assume an SPS distribution that exhibits a monotonic decrease in its $k$th-order correlation functions $g^{(k+1)}(0) \leq g^{(k)}(0)$ for all $k \geq 2$, which is experimentally verifiable. Expressing each emission probability $p^{(\mathrm{MD})}_{k}$ in terms of its associated $g^{(k)}(0)$, the upper bound of the multiphoton probability is given by
\begin{align*}
 p_{\mathrm{mp}}^{(\mathrm{MD})} &= p_2^{(\mathrm{MD})} + \sum_{k=3}^{\infty} p_k^{(\mathrm{MD})} \\
 &= \frac{g^{(2)}(0) \braket{n}^2}{2} + \sum_{k=3}^{\infty}  \frac{k-1}{k} (-1)^k  g^{(k)}(0) \braket{n}^k \\
 & \leq \overline{p}_{\mathrm{mp}}^{(\mathrm{MD})} = \overline{p}_{\mathrm{mp}} + g^{(2)}(0)  \underbrace{\sum_{k=3}^{\infty}  \frac{k-1}{k} (-1)^k \braket{n}^k}_{< 0},
\end{align*}
hence $\overline{p}_{\mathrm{mp}}^{(\mathrm{MD})} \leq \overline{p}_{\mathrm{mp}}$. Note that in any truncated Hilbert space, making the summation finite, this inequality holds. Though counterintuitive, we conclude that any distribution with states higher than two-photons decreases the overall multiphoton probability. Consequently, to simulate Bob's counts, we implement the distribution $\{p_k^{(\mathrm{MD})}\}_{k = 0,1,2}$ with $p_{k > 2}^{(\mathrm{MD})} = 0$, since it satisfies the equality $p_{\mathrm{mp}}^{(\mathrm{MD})} = p_{2}^{(\mathrm{MD})} = \overline{p}_{\mathrm{mp}}$. Thus, Alice's emission probabilities read as
\begin{align}
        p_2^{(\mathrm{MD})} &= \frac{g^{(2)}(0) \braket{n}^2}{2} \\
        \label{p1MD}
        p_1^{(\mathrm{MD})} &= \braket{n} - 2p_2^{(\mathrm{MD})} \\
        \label{p0MD}
        p_0^{(\mathrm{MD})} &= 1 - p_2^{(\mathrm{MD})} - p_1^{(\mathrm{MD})}.
\end{align}

Furthermore, pre-attenuating Alice's source increases the maximum tolerable channel loss, defined as the highest loss that generates a positive key. We define this pre-attenuation by a transmissivity value $\eta_{\mathrm{tr}}$, representing the fraction of signal that goes through the attenuator to the quantum channel. This transmissivity reduces the multiphoton probability quadratically $\overline{p}_m^{(\mathrm{att})} = g^{(2)}(0) \braket{n}^2 \eta_{\mathrm{tr}}^2 / 2$, while decreasing the click probability at Bob's detectors linearly at a first-order Taylor approximation, $p_{\mathrm{click}} = \sum_{n=0}^{\infty} p_k^{(\mathrm{MD})} \left[ 1 - (1 - p_{\mathrm{dc}}) \left(1 - \eta_{\mathrm{tr}} \eta_{\mathrm{ch}} \eta_{\mathrm{det}}  \right)^n\right] \approx p_{\mathrm{dc}} + \left( 1 - p_{\mathrm{dc}} \right) \eta_{\mathrm{tr}} \eta_{\mathrm{ch}} \eta_{\mathrm{det}} \braket{n}$, where $\eta_{\mathrm{ch}}$ and $\eta_{\mathrm{det}}$ are the channel and detector efficiencies, respectively, and $p_{\mathrm{dc}}$ is the dark count probability. Although we compute the exact expression in our model, this approximation is valid as states with two or more photons do not dominate the source photon emission. Therefore, the multiphoton probability decreases more rapidly than the click probability on Bob's side, widening the range of tolerable losses. This enhances the key generation at the high-loss regime where $\overline{p}_m^{(\mathrm{att})}$ dominates, obtaining the highest possible secure key by optimising $\eta_{\mathrm{tr}}$ for each channel loss. Additionally, the basis bias $p_X$, the probability of choosing the $X$ basis by either party, is also optimised. Note that $p_Z = 1 - p_X$ for the parameter estimation basis. Setting an unequal bias has been reported as an exceptional strategy to increase the key generation~\cite{wei2013decoy, yu2016reexamination}.

\textit{Secure key length estimation: asymptotic and finite-key analysis.} We distinguish two scenarios for the secure key length (SKL) estimation: the asymptotic and finite-key analysis. In the asymptotic limit, we assume the experiment runs for an infinity time duration, resulting in a sufficiently large number of detection events in the key generation basis, allowing us to consider $p_X \rightarrow 1$, and a negligible phase error rate, $p_Z \rightarrow 0$. Consequently, the count rates converge to their underlying true expectation values. Asymptotic secure key rates were already proposed long ago such as the Devetak-Winter bound~\cite{devetak2005distillation}. However, in this work, we compute the asymptotic key rates by simply setting a sufficiently large block size, which yields identical outcomes to the asymptotic formula.

However, in a real experiment, the obtained statistics are finite and subject to fluctuations from their expected outcomes. Consequently, studies were proposed to account for the effect of finite statistics~\cite{hasegawa2007security}. Here, this issue is addressed as follows: with generality, we define the number of multiphoton states received by Bob as a finite set of independent Bernoulli random variable $\{X_1^{B}, X_2^B, \cdots, X_{N_{\mathrm{S}}}^B\}$ with two possible outcomes $\{0,1\}$. Its observed value is $X^B \equiv \sum_{i=1}^{N_{\mathrm{S}}} X_i^B$ that satisfy $\mathrm{Pr} \left( X_i^B = 1 \right) = P_{i, \mathrm{click} | m} \, P_{i, m}^{(att)}$ for fixed fixed $m \geq 2$, i.e. the product of the conditional probability of a click when a multiphoton is sent after Alice's pre-attenuation and the probability of sending a multiphoton. Note that each random variable denoted by the subscript $i$ is associated with a multiphoton state with a fixed number of photons $m \geq 2$, which may change for each variable, hence their probabilities too. To prevent an overestimation of the secure events, we assume the worst scenario where every sent multiphoton, which is untrustworthy, causes a click on Bob's apparatus, i.e. $P_{i, \mathrm{click} | m} = 1$. Therefore, an expected value of $X^{B}$ is expressed as $X^{B \, *} = \sum_{i=1}^{N_{\mathrm{S}}} P_{i, m}^{(att)}$, where $N_{\mathrm{S}} = R_{\mathrm{rate}} t$ is the number of sent signals by Alice that forms the finite block, defined by the acquisition time $t$, i.e. the time the experiment is run, and the source repetition rate $R_{\mathrm{rate}}$. However, since we lack access to the true attenuated photon emission probabilities of the SPS $P_{i, m}^{(att)}$, we bound them as $X^{B \, *} = \sum_{i=1}^{N_{\mathrm{S}}} P_{i, m}^{(att)} \leq \sum_{i=1}^{N_{\mathrm{S}}} \overline{p}_{\mathrm{mp}}^{(att)} = N_{\mathrm{S}} \overline{p}_{\mathrm{mp}}^{(att)}$. Note that even if the legitimate parties had access to them from perfect SPS characterisation, they would not know how many photons Alice sends in each state. After sifting, the expected number of multiphoton events received by Bob in the key generation basis is $N_{\mathrm{R,mp}}^{X \; *} = p_X^2 X^{B \, *} \leq  N_{\mathrm{S}} p_X^2 \overline{p}_{\mathrm{mp}}^{(att)} = \overline{N}_{\mathrm{R,mp}}^{X \; *}$.

In a real QKD experiment, Bob observes clicks from his detector and unfortunately, even if Bob measures the multiphoton events using a photon number resolving detector, his results cannot be trusted due to potential PNS attacks by Eve. Therefore, we need a consistent method that for a given expected value $N_{\mathrm{R,mp}}^{X \; *}$ of a data block, derives an upper bound of the observed value $\overline{N}_{\mathrm{R,mp}}^{X}$, whose tail probability is bounded with a parameter estimation failure probability as $\mathrm{Pr} 
\left[ N_{\mathrm{R,mp}}^{X} \geq \left( 1 + \Delta^U \right) N_{\mathrm{R,mp}}^{X \; *} \right] \leq \varepsilon_{\mathrm{PE}} = 2 \varepsilon_{\mathrm{sec}} / 3$.

Several methods were proposed to account for these statistical fluctuations in finite blocks for decoy WCP methods, such as the Gaussian analysis method~\cite{ma2005practical}, the Hoeffding inequality~\cite{lim2014concise} and the multiplicative Chernoff bound~\cite{curty2014finite}. However, an improved analytical Chernoff bound has reported tighter finite-key bounds, enhancing the key generation~\cite{yin2020tight}. Here, for our 2-decoy WCP protocol, we employ this updated Chernoff bound with the decoy method of~\cite{curty2014finite}, as presented in the analysis of~\cite{sidhu2022finite} but for a fibre link instead of a satellite-to-ground link. Our non-decoy SPS protocol of ~\cite{morrison2023single} has already shown massive key rate enhancements using the same updated Chernoff bound as the decoy WCP studies, compared to previous mathematical deviations applied to protocol probabilities~\cite{cai2009finite}. Thus, we apply their Chernoff bound to estimate the upper bound of the observed value $\overline{N}_{\mathrm{R,mp}}^{X} = \overline{N}_{\mathrm{R,mp}}^{X \; *} + \Delta^U$ with $\Delta^U = \left( \beta + \sqrt{8\beta \overline{N}_{\mathrm{R,mp}}^{X \; *} + \beta^2} \right) / 2 \overline{N}_{\mathrm{R,mp}}^{X \; *}$ where $\beta = - \ln \varepsilon_{\mathrm{PE}}$. The lower bound of the observed non-multiphoton events in the key generation basis is $\underline{N}_{\mathrm{R,nmp}}^{X} = N_{\mathrm{R}}^{X} - \overline{N}_{\mathrm{R,mp}}^{X}$, where $N_{\mathrm{R}}^{X} = N_{\mathrm{S}} p_X^2 p_{\mathrm{click}}$ is the observed number of detection events in the $X$ basis by Bob. Likewise, $N_{\mathrm{R}}^{Z} = N_{\mathrm{S}} p_Z^2 p_{\mathrm{click}}$ and $\underline{N}_{\mathrm{R,nmp}}^{Z} = N_{\mathrm{R}}^{Z} - \overline{N}_{\mathrm{R,mp}}^{Z}$ are calculated for the parameter estimation basis. Note that here the security condition against PNS attacks $p_{\mathrm{click}} > \overline{p}_{\mathrm{mp}}^{(att)}$ is imposed because if it is not met, $\underline{N}_{\mathrm{R,nmp}}^{X}$ is negative, resulting in no shared key, see eq. (\ref{eq:fkr}). The number of errors is determined as $m_X = N_{\mathrm{S}} p_X^2 p_{\mathrm{err}}$ and $m_Z = N_{\mathrm{S}} p_Z^2 p_{\mathrm{err}}$ for each basis, where given the error probability due to misalignment of the set-up $p_{\mathrm{mis}}$ the error probability reads as
\begin{equation}
\begin{split}
    p_{\mathrm{err}} &= \frac{p_0 p_{\mathrm{dc}}}{2}  \\
    &+ \sum_{n=1}^{\infty} p_n \left[ 1 - (1 - p_{\mathrm{dc}}) (1 - \eta_{ch} \eta_{det} \eta_{att})^n\right]  p_{\mathrm{mis}} .
    \end{split}
\end{equation}
Note $m_X$ is not publicly revealed and is only used to estimate the number of bits needed to perform error correction.

Subsequently, we blind ourselves to this simulation model and work with the observed outcomes to estimate the secure key length (SKL). We calculate the SKL for each finite block defined by the number of sent signals $N_{\mathrm{S}}$. The main steps of the efficient BB84 protocol proceed as follows: Alice sends $N_{\mathrm{S}}$ states from her pre-attenuated SPS and Bob measures them, obtaining $N_R$ detection events. This block is split into three sub-blocks: the discarded events due to sifting $2p_X \left( 1 - p_X\right)$, the events used for key generation $p_X^2$ and the events used for parameter estimation $\left(1-p_X\right)^2$. Finally, the extracted SKL from the sifted key is given by~\cite{morrison2023single}
\begin{equation}
    \label{eq:fkr}
    \ell_{\mathrm{SPS}} = \underline{N}_{\mathrm{R,nmp}}^X \left[1 - H\left(\overline{\phi}^X\right)\right] -\lambda_{\mathrm{EC}} - 2\log_2 \frac{3}{\varepsilon_{\mathrm{sec}}} - \log_2 \frac{2}{\varepsilon_{\mathrm{cor}}} ,
\end{equation}
where $1 - H\left(\overline{\phi}^X\right)$ represents the information leaked to Eve, bounded by the binary entropy $H(e) = - e \log_2 e - (1-e) \log_2 (1-e)$. We set a phase error rate threshold of $\phi_{\mathrm{th}}^X = 0.11$ (11 $\%$)~\cite{shor2000simple}. Thus, the legitimate parties abort the protocol if $\overline{\phi}^X \geq \phi_{\mathrm{th}}^X$, assuming the presence of an eavesdropper. The parties publicly announce their errors in the parameter estimation basis $m_Z$ to estimate the phase error rate caused by non-multiphotons in the key generation basis $\phi_X = m_Z / \underline{N}_{\mathrm{R,nmp}}^Z$, which is upper-bounded for the $N_R^X$ sample, that is not revealed, as $\overline{\phi}^X = \phi_X + \gamma^U \left( N_{\mathrm{R}}^X, N_{\mathrm{R}}^Z, \phi_X, \varepsilon_{\mathrm{PE}} \right)$ using the $\gamma^U$ function of~\cite{yin2020tight} for the random sampling without replacement problem. $\lambda_{\mathrm{EC}}$ accounts for the number of bits used in the error correction code and we use the improved approximation of~\cite{tomamichel2017fundamental}. However, no practical code has reached a value below the Shannon limit of $1.16 N_{\mathrm{R}}^X H \left( e_X \right)$ where $e_X = m_X / N_{\mathrm{R}}^X$ is the quantum bit error rate. Therefore, if $\lambda_{EC} / N_{\mathrm{R}}^X H(e_X) < 1.16$, we recover the Shannon limit to estimate the bits used in error correction. Finally, the last two terms represent the secrecy and correctness parameters, $\varepsilon_{\mathrm{sec}}$ and $\varepsilon_{\mathrm{cor}}$ respectively, which ensures the protocol is $\varepsilon_{\mathrm{QKD}} = \varepsilon_{\mathrm{sec}} + \varepsilon_{\mathrm{cor}}$ secure in the composable security framework~\cite{renner2008security}. 

Since we lump together the vacuum and single-photon detection events into the non-multiphoton estimation, one may think adding a vacuum distribution allows us to estimate them separately and more accurately, hence enhancing the key rate. Here we show this is not the case. We consider two new scenarios: first, Bob knows exactly the vacuum contribution, $N_{\mathrm{R,0}}^{X} = N_{\mathrm{S}} p_X^2 p_0^{(\mathrm{MD})} p_{\mathrm{click}}$; second, Bob estimates a lower bound of the vacuum events. For the latter, using the mean photon number definition, we estimate the lower bound of the vacuum emission probability as $P_0 \geq \underline{p}_0 = 1 - \braket{n}$ and then the vacuum events as $\underline{N}_{\mathrm{R,0}}^{X} = N_{\mathrm{S}} p_X^2 \underline{p}_0 p_{\mathrm{click}}$.

\begin{figure}[htb!]
    \centering
    \includegraphics[width=0.95\columnwidth]{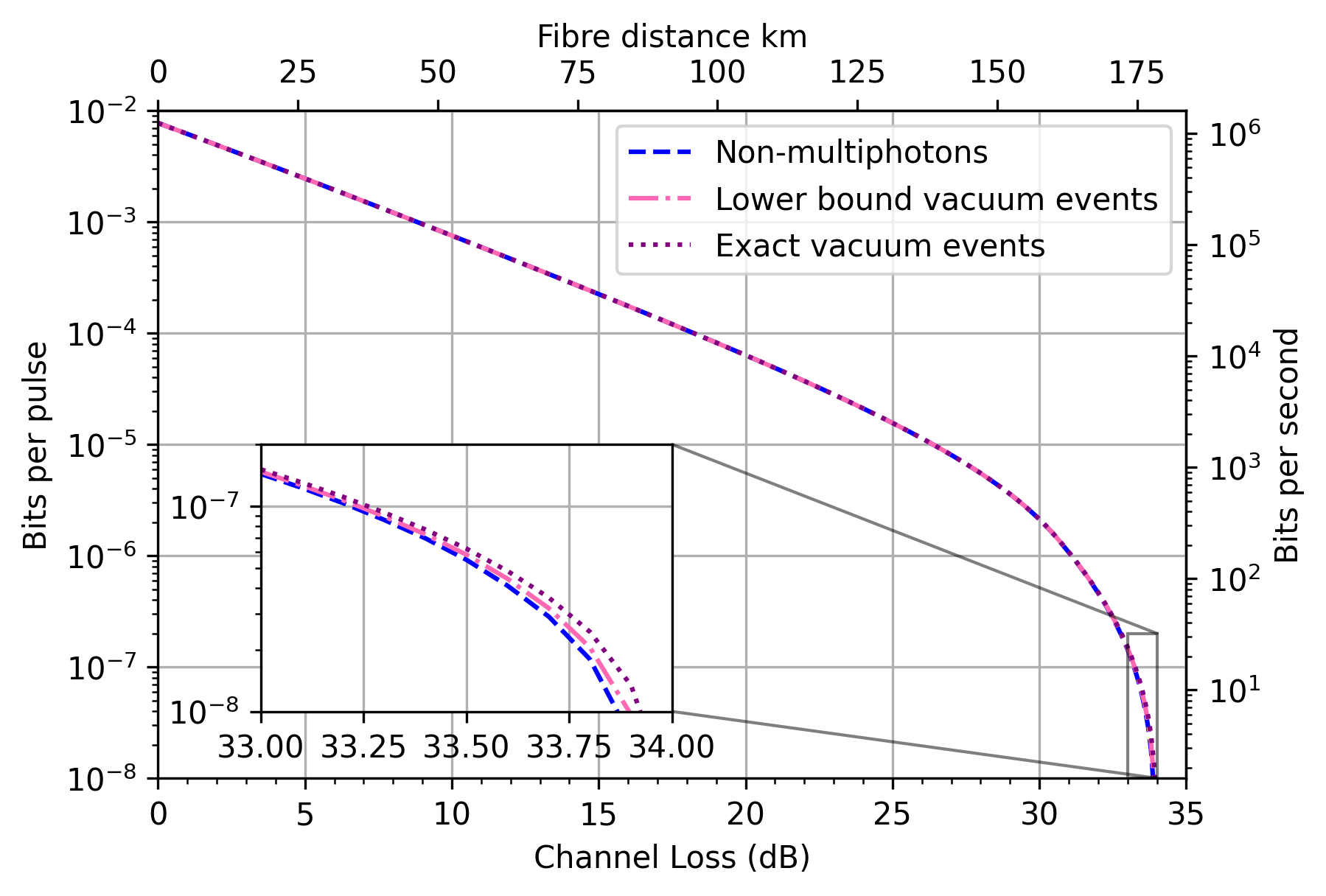}
    \caption{The secure key of the non-decoy SPS protocol as a function of the channel loss (fibre distance) in a semi-log scale with optimised $p_X$ and $\eta_{tr}$. The $g^{(2)}(0) = 0.036$ and the mean photon number $\braket{n} = 0.0142$ are fixed for $1$ minute of acquisition time. The dashed blue curve represents the SKL using the non-multiphoton estimation, and the other two curves estimate separately the vacuum and single-photon events, the green dotted curve uses lower bound on the vacuum events $\underline{N}_{\mathrm{R,0}}^{X}$ and the dashed purple curve uses the exact number of vacuum events $N_{\mathrm{R,0}}^{X}$.}
    \label{vacuum_yield_1}
\end{figure}

The key enhancement showed by these two scenarios with a vacuum decoy state goes unnoticed for all channel losses. In particular, in the high-loss regime, even assuming Bob has complete knowledge of the vacuum contribution, the maximum tolerable loss increases by only $0.05$ dB for one minute of acquisition time, see Figure \ref{vacuum_yield_1}. Therefore, due to its modest key rate increase, the extra experimental endeavour, considering the usual difficulty of creating a perfect decoy vacuum state in practice~\cite{rosenberg2007long,dixon2008gigahertz}, and the need for additional detector characterisation to estimate the vacuum contribution, is not worth considering. Thus, we also show the lower bound of the non-multiphoton events is not underestimated compared to estimating the events separately. As a result, we take this conservative approach and assume that Eve gains the same amount of information from the vacuum states as from the single-photon states.

As mentioned above, we also employ the updated Chernoff bound for a 2-decoy protocol with WCPs. Thus, we use the finite-key analysis and parameter optimisation of~\cite{sidhu2022finite}, considering one decoy state with a lower intensity than the signal state and a vacuum decoy state, whose secret key length is given by
\begin{equation}
\label{eqn:skl_lim_result}
\begin{split}
    \ell_{\mathrm{WCP}} = & \, \underbar{N}_{\mathrm{R,0}}^X  + \underbar{N}_{\mathrm{R,1}}^X \left[ 1 - H(\bar{\phi}^X) \right] - \lambda_{\mathrm{EC}} \\
    & - 6 \log_2 \frac{21}{\varepsilon_{\mathrm{sec}}} - \log_2 \frac{2}{\varepsilon_{\mathrm{cor}}},
\end{split}
\end{equation}
where $\underbar{N}_{\mathrm{R,0}}^X$, $\underbar{N}_{\mathrm{R,1}}^X$ are the lower bounds of the vacuum and single-photon events respectively. Note that the rest of the parameters follow the same criteria as in the non-decoy SPS protocol. Finally, the secure key rate (SKR) is defined as $r_{\mathrm{SPS}} = \ell_{\mathrm{SPS}} / N_{\mathrm{S}}$ and $ r_{\mathrm{WCP}} = \ell_{\mathrm{WCP}} / N_{\mathrm{S}}$ for non-decoy SPS and 2-decoy WCP, respectively.

\begin{figure}[htb!]
    \centering
    \includegraphics[width=1.0\columnwidth]{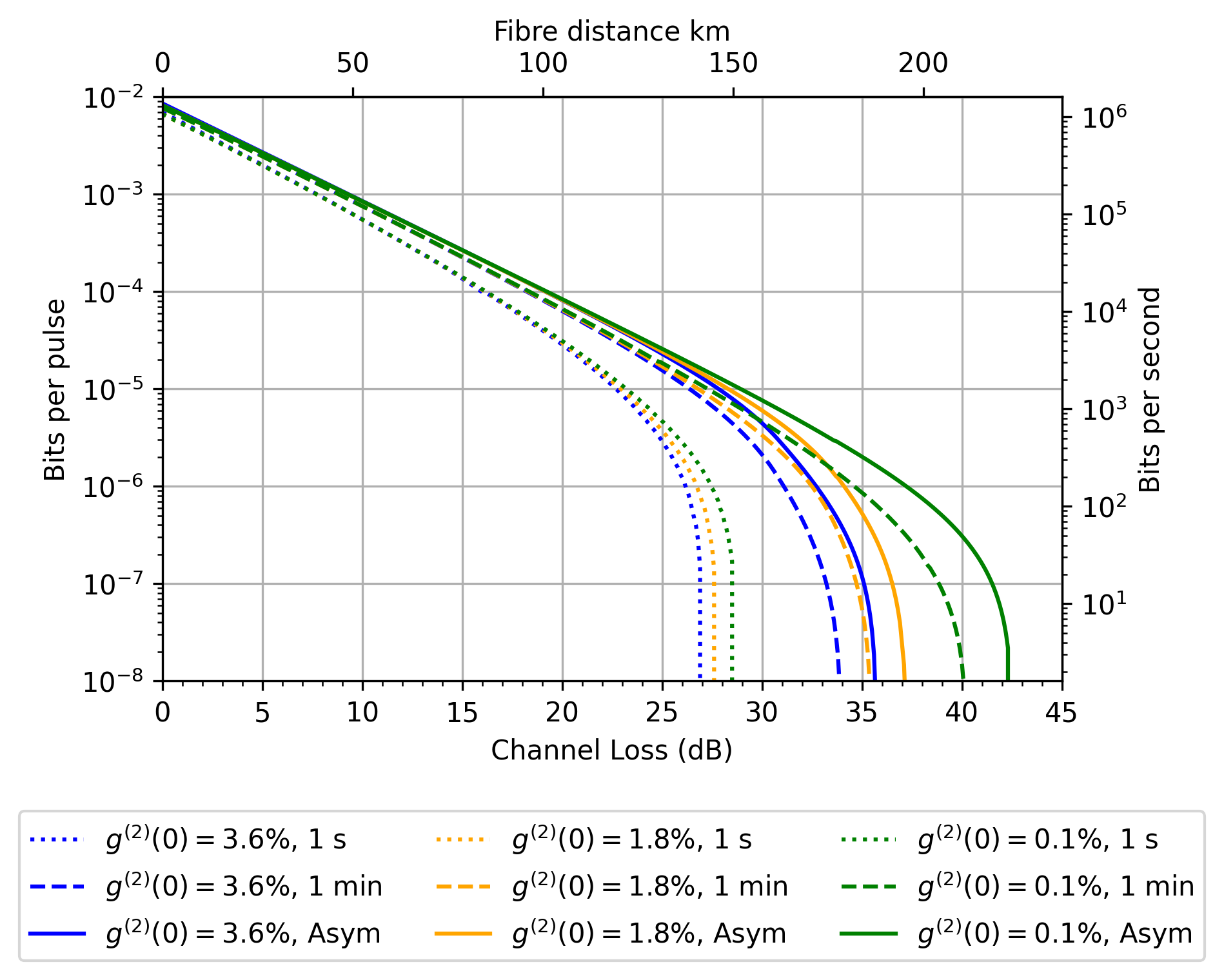}
    \caption{Comparison of the secure key in a semi-log scale for the non-decoy SPS protocol with several second-order correlation functions $g^{(2)}(0)$ and block sizes, optimising the basis bias $p_X$ and the transmissivity $\eta_{tr}$ associated with Alice's source pre-attenuation for each channel loss. The SKR $r_{\mathrm{SPS}}$ and SKL $\ell_{\mathrm{SPS}}$, eq. (\ref{eq:fkr}), are represented by the bits per pulse and the bits per second, respectively. The $g^{(2)}(0) = 3.6\%$ (blue curves) corresponds to the quantum dot used in the QKD analysis of~\cite{morrison2023single}. The $g^{(2)}(0) = 1.8\%$ (orange curves) is based on~\cite{rakhlin2023demultiplexed} and $g^{(2)}(0) = 0.1\%$ (green curves) is considered an optimistic case. The mean photon number is fixed to $\braket{n} = 0.0142$, which corresponds to the quantum dot of~\cite{morrison2023single}. We consider two acquisition times (number of sent signals) of 1 second ($1.607 \cdot 10^{8}$), dotted curves, and 1 minute ($9.642 \cdot 10^{9}$), dashed curves, plus the asymptotic limit, solid curves.}
    \label{SKRvsLossg2}
\end{figure}

\begin{figure}[htb!]
    \centering
    \includegraphics[width=1.0\columnwidth]{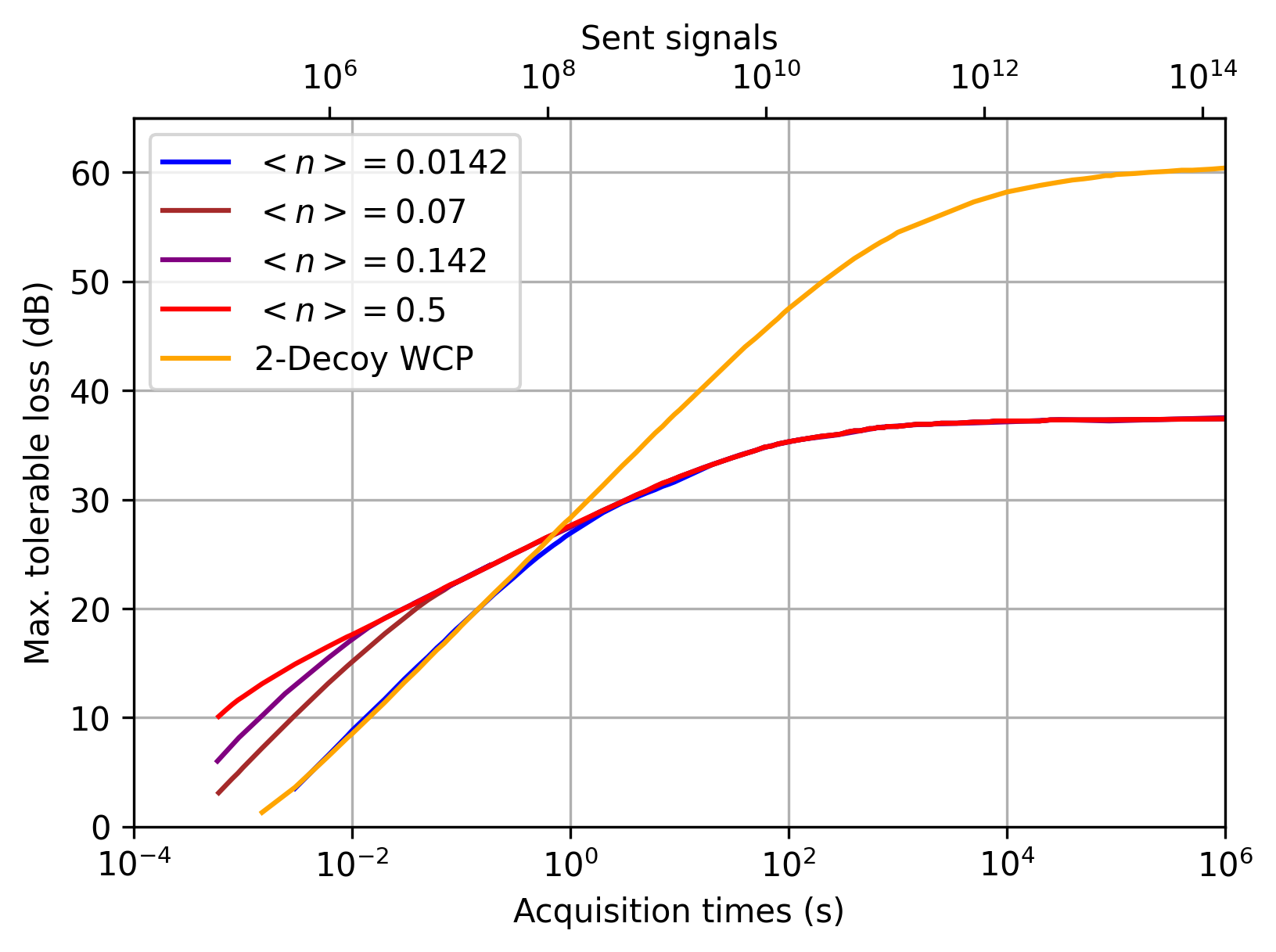}
    \caption{Comparison of the maximum tolerable channel loss as a function of the acquisition time in a semi-log scale for the non-decoy SPS protocol with several mean photon numbers $\braket{n}$ and the 2-decoy WCP protocol (orange curve), optimising the free parameters in both protocols. The maximum loss is displayed as a function of the block size represented by the acquisition time or the number of sent signals by Alice. Here the second-order correlation function is fixed to $g^{(2)}(0) = 3.6\%$.}
    \label{MaxTolLoss}
\end{figure}

\textit{Discussion.} Here, we analyse the impact of different SPS characteristics on the secure key and the maximum tolerable loss for several finite blocks. The block size used to extract the secure key is defined by the number of signals sent by Alice which depends on the acquisition time of the experiment. We compare the results of the non-decoy SPS protocol with the 2-decoy WCP protocol. The fixed QKD parameters to all protocols and figures are shown in Table \ref{Table_parameters}.

\begin{table}[htb!]
    \begin{adjustbox}{width=1.0\columnwidth}
    \centering
    \begin{tabular}{l*{3}{l}}
        \hline
        \textbf{Description}\vspace{2pt} & \textbf{Parameter}\vspace{2pt} & \textbf{Value} \vspace{2pt} \\
        \hline
        Source repetition rate & $R_{\mathrm{rate}}$ & $160.7$ MHz \\
        Misalignment probability & $p_{\mathrm{mis}}$ & $0.003$ \\
        Dark count probability & $p_{\mathrm{dc}}$ & $3.67\times10^{-8}$ \\
        Detector efficiency & $\eta_{\mathrm{det}}$ & $0.6525$ \\
        Fibre loss & $l$ & $0.1904$ dB/km \\
        Secrecy failure probability & $\varepsilon_{\mathrm{sec}}$ & $10^{-10}$ \\
        Correctness failure probability & $\varepsilon_{\mathrm{cor}}$ & $10^{-15}$ \\
        \hline
    \end{tabular}
    \end{adjustbox}
    \caption{Baseline QKD protocol parameters based on the experiment of the quantum dot ~\cite{morrison2023single}.}
    \label{Table_parameters}
\end{table}

In Figure \ref{SKRvsLossg2}, we show the impact of $g^{(2)}(0)$ on the secure key performance for channel losses (optical fibre distances) in the non-decoy SPS protocol. As expected, fixing the dark count probability means that the $g^{(2)}(0)$ value determines the drop-off of the secure key curves, hence the maximum tolerable loss. For a one-second time block (dotted curves), the improvement in the maximum tolerable loss is approximately $1$ dB between the three $g^{(2)}(0)$ values, which is modest. However, for one minute of acquisition time (dashed curves), the difference in the range of tolerable losses among the $g^{(2)}(0)$ values exhibits a considerable increment since the secure key performance is already remarkably close to the asymptotic regime. The asymptotic key of $g^{(2)}(0) = 3.6\%$ (solid blue curve) is approached at one minute of acquisition time by halving the $g^{(2)}(0)$ (dashed orange line). It is worth mentioning that one hour of acquisition time is needed for $g^{(2)}(0) = 3.6\%$ to approach its asymptotic key rate. Therefore, halving the $g^{(2)}(0)$ from $3.6\%$ to $1.8\%$ produces similar key rates but for massively different block sizes, resulting in a reduction of approximately $98\%$ in the acquisition time.

\begin{figure}[htb!]
    \centering
    \includegraphics[width=1.0\columnwidth]{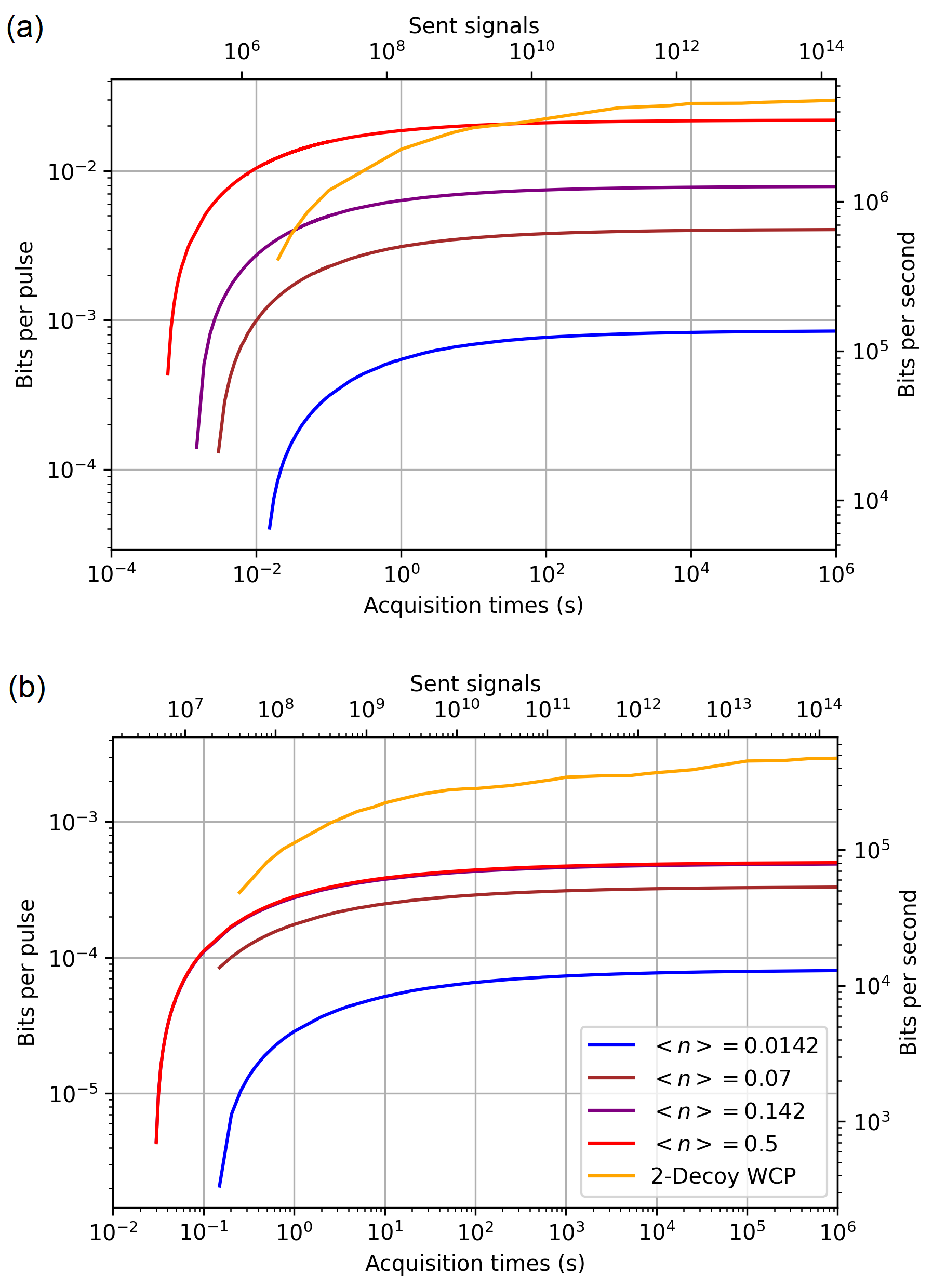}
    \caption{Secure key as a function of the acquisition times (sent signals) in log-log scale for (a) $10$ dB and (b) $20$ dB of channel loss for the optimised non-decoy SPS protocol with various mean photon numbers $\braket{n}$ and the optimised 2-decoy WCP method (orange curve). Here the $g^{(2)}(0) = 0.036$.}
    \label{SKR10and20dB}
\end{figure}

In Figure \ref{MaxTolLoss}, we show the maximum tolerable channel loss varying the mean photon number $\braket{n}$ while fixing $g^{(2)}(0)$. A higher $\braket{n}$ represents a lower vacuum emission and higher single-photon and multiphoton emissions. Therefore, the higher $\braket{n}$ is, the lower acquisition time (number of sent signals) is required for the pre-attenuation of Alice's source to kick in. The SPS with $\braket{n} = 0.5$ (red curve) introduces the pre-attenuation even for the smallest acquisition time. This indicates that any other source with a mean photon number $\braket{n} > 0.5$ will not be able to cause a rise in the maximum tolerable loss, as the distribution will be pre-attenuated by Alice anyway. The point at which each curve converges to the curve with $\braket{n} = 0.5$ represents its first acquisition time in which the pre-attenuation is introduced. For acquisition times above one second, all the key curves reach the secure key of $\braket{n} = 0.5$ and their asymptotic limit is achieved above 100 seconds where the maximum loss is constant despite the rise of time. In particular, for the SPS of \cite{morrison2023single} (blue curve), there is an increase between $0.1$ and $0.2$ dB on the maximum tolerable loss for acquisition times within the interval of $\left[0.01, 0.1\right]$ seconds. A greater mean photon number than $\braket{n} > 0.0142$ is needed to tolerate higher losses than the 2-decoy state protocol for times below $0.62$ seconds ($10^8$ sent signals). Consequently, an SPS with $\braket{n} > 0.0142$ would extend the range of tolerable losses compared to a decoy WCP system with a maximum tolerable channel loss of $25$ dB or below. Outside this range, the superiority of the 2-decoy WCP is evident. However, it is worth mentioning the SPS with the highest mean photon number, $\braket{n} = 0.5$, for extremely short acquisition times of $0.01$ seconds, the SPS protocol tolerates up to $9$ dB ($47$ km) more channel loss (fibre distance) than the WCP protocol. 

We also fix the channel loss and show the secure key rate versus time blocks. Note that for the fixed losses in Figure \ref{SKR10and20dB}, a higher mean photon number than $\braket{n} = 0.5$ (red curve) would produce the same key rate, since it would be pre-attenuated and $g^{(2)}(0)$ does not dominate in this low-loss regime. Therefore, $\braket{n} = 0.5$ shows the highest possible secure key generation for the SPS protocol at these channel losses. This principle is illustrated in Figure \ref{SKR10and20dB} (b) with 20 dB of channel loss, where the key results of $\braket{n} = 0.5$ (red curve) overlap with the secure key curve of $\braket{n} = 0.142$ (purple curve). In the regime when the block size tends to the asymptotic limit, the highest key rate (red curve) scales by a factor of $5$ compared to the lowest $\braket{n} = 0.0142$. Furthermore, non-decoy SPS only shows an advantage over the 2-decoy WCP at 20 dB loss for extremely small acquisition times, where the 2-decoy method is unable to generate key. In Figure \ref{SKR10and20dB} (a) with a 10 dB of channel loss, the non-decoy SPS of $\braket{n} = 0.142$ (purple line) and $\braket{n} = 0.5$ (red line) outperform the secure key generation of the 2-decoy WCP (orange curve) for block sizes approximately below $0.03$ seconds ($4.8 \cdot 10^6$ sent signals) and $29$ seconds ($4.66 \cdot 10^9$ sent signals), respectively.

\textit{Conclusions.} We demonstrate that the non-decoy SPS may enhance the key rate performance of the 2-decoy WCP for short acquisition times (small number of sent signals) in the low-loss regime. Additionally, the range of acquisition times where the non-decoy SPS surpasses the 2-decoy WCP can be potentially wider for channel losses below $10$ dB ($52.5$ km of fibre). Within the same regime, at least a mean photon number as the quantum dot in~\cite{morrison2023single} is required to tolerate higher losses over the 2-decoy WCP. Complementary SPS characteristics for both improvements include a $g^{(2)}(0) \leq 3.6\%$ and a source repetition rate of $R_{\mathrm{rate}} \geq 160.7$ MHz. Through our theoretical key estimates, we present SPSs as a valid quantum source for short-range QKD, performing an efficient BB84 protocol with one unique source at Alice's side, thus avoiding experimental complexity associated with decoy methods. Nevertheless, further research is necessary to establish a fair comparison between WCPs and SPSs within the decoy method framework. Concretely, a comparison to decoy analysis using a Fock basis notation that aligns with the SPS approach outline here would be beneficial~\cite{wang2009decoy}.

\textit{Acknowledgments.} The authors express gratitude to C. Morrison and A. Fedrizzi for their insightful discussions, which not only contributed to our previous experimental paper but also inspired the expansion of the analysis presented in this work. R.G.P. acknowledges support from the EPSRC Research Excellence Award
(REA) Studentship. R.G.P. and D.K.L.O. are supported by the EPSRC International Network in Space Quantum Technologies (EP/W027011/1) and the Quantum Technology Hub in Quantum Communication (EP/T001011/1). D.K.L.O. is supported by the EPSRC Researcher in Residence programme at the Satellite Applications Catapult (EP/T517288/1). J.J. is supported by QuantIC, the EPSRC Quantum Technology Hub in Quantum Imaging (EP/T00097X/1).

\bibliography{NonDecoyPaper}

\begin{thebibliography}{37}%
\makeatletter
\providecommand \@ifxundefined [1]{%
 \@ifx{#1\undefined}
}%
\providecommand \@ifnum [1]{%
 \ifnum #1\expandafter \@firstoftwo
 \else \expandafter \@secondoftwo
 \fi
}%
\providecommand \@ifx [1]{%
 \ifx #1\expandafter \@firstoftwo
 \else \expandafter \@secondoftwo
 \fi
}%
\providecommand \natexlab [1]{#1}%
\providecommand \enquote  [1]{``#1''}%
\providecommand \bibnamefont  [1]{#1}%
\providecommand \bibfnamefont [1]{#1}%
\providecommand \citenamefont [1]{#1}%
\providecommand \href@noop [0]{\@secondoftwo}%
\providecommand \href [0]{\begingroup \@sanitize@url \@href}%
\providecommand \@href[1]{\@@startlink{#1}\@@href}%
\providecommand \@@href[1]{\endgroup#1\@@endlink}%
\providecommand \@sanitize@url [0]{\catcode `\\12\catcode `\$12\catcode `\&12\catcode `\#12\catcode `\^12\catcode `\_12\catcode `\%12\relax}%
\providecommand \@@startlink[1]{}%
\providecommand \@@endlink[0]{}%
\providecommand \url  [0]{\begingroup\@sanitize@url \@url }%
\providecommand \@url [1]{\endgroup\@href {#1}{\urlprefix }}%
\providecommand \urlprefix  [0]{URL }%
\providecommand \Eprint [0]{\href }%
\providecommand \doibase [0]{https://doi.org/}%
\providecommand \selectlanguage [0]{\@gobble}%
\providecommand \bibinfo  [0]{\@secondoftwo}%
\providecommand \bibfield  [0]{\@secondoftwo}%
\providecommand \translation [1]{[#1]}%
\providecommand \BibitemOpen [0]{}%
\providecommand \bibitemStop [0]{}%
\providecommand \bibitemNoStop [0]{.\EOS\space}%
\providecommand \EOS [0]{\spacefactor3000\relax}%
\providecommand \BibitemShut  [1]{\csname bibitem#1\endcsname}%
\let\auto@bib@innerbib\@empty
\bibitem [{\citenamefont {Hellman}(2002)}]{hellman2002overview}%
  \BibitemOpen
  \bibfield  {author} {\bibinfo {author} {\bibfnamefont {M.~E.}\ \bibnamefont {Hellman}},\ }\bibfield  {title} {\bibinfo {title} {An overview of public key cryptography},\ }\href@noop {} {\bibfield  {journal} {\bibinfo  {journal} {IEEE Communications Magazine}\ }\textbf {\bibinfo {volume} {40}},\ \bibinfo {pages} {42} (\bibinfo {year} {2002})}\BibitemShut {NoStop}%
\bibitem [{\citenamefont {Grollmann}\ and\ \citenamefont {Selman}(1988)}]{grollmann1988complexity}%
  \BibitemOpen
  \bibfield  {author} {\bibinfo {author} {\bibfnamefont {J.}~\bibnamefont {Grollmann}}\ and\ \bibinfo {author} {\bibfnamefont {A.~L.}\ \bibnamefont {Selman}},\ }\bibfield  {title} {\bibinfo {title} {Complexity measures for public-key cryptosystems},\ }\href@noop {} {\bibfield  {journal} {\bibinfo  {journal} {SIAM Journal on Computing}\ }\textbf {\bibinfo {volume} {17}},\ \bibinfo {pages} {309} (\bibinfo {year} {1988})}\BibitemShut {NoStop}%
\bibitem [{\citenamefont {Bennett}\ \emph {et~al.}(1992)\citenamefont {Bennett}, \citenamefont {Brassard},\ and\ \citenamefont {Ekert}}]{bennett1992quantum}%
  \BibitemOpen
  \bibfield  {author} {\bibinfo {author} {\bibfnamefont {C.~H.}\ \bibnamefont {Bennett}}, \bibinfo {author} {\bibfnamefont {G.}~\bibnamefont {Brassard}},\ and\ \bibinfo {author} {\bibfnamefont {A.~K.}\ \bibnamefont {Ekert}},\ }\bibfield  {title} {\bibinfo {title} {Quantum cryptography},\ }\href@noop {} {\bibfield  {journal} {\bibinfo  {journal} {Scientific American}\ }\textbf {\bibinfo {volume} {267}},\ \bibinfo {pages} {50} (\bibinfo {year} {1992})}\BibitemShut {NoStop}%
\bibitem [{\citenamefont {Bennett}\ and\ \citenamefont {Brassard}(2014)}]{bennett2014quantum}%
  \BibitemOpen
  \bibfield  {author} {\bibinfo {author} {\bibfnamefont {C.~H.}\ \bibnamefont {Bennett}}\ and\ \bibinfo {author} {\bibfnamefont {G.}~\bibnamefont {Brassard}},\ }\bibfield  {title} {\bibinfo {title} {Quantum cryptography: Public key distribution and coin tossing},\ }\href@noop {} {\bibfield  {journal} {\bibinfo  {journal} {Theoretical computer science}\ }\textbf {\bibinfo {volume} {560}},\ \bibinfo {pages} {7} (\bibinfo {year} {2014})}\BibitemShut {NoStop}%
\bibitem [{\citenamefont {Wang}\ \emph {et~al.}(2019)\citenamefont {Wang}, \citenamefont {He}, \citenamefont {Chung}, \citenamefont {Hu}, \citenamefont {Yu}, \citenamefont {Chen}, \citenamefont {Ding}, \citenamefont {Chen}, \citenamefont {Qin}, \citenamefont {Yang} \emph {et~al.}}]{wang2019towards}%
  \BibitemOpen
  \bibfield  {author} {\bibinfo {author} {\bibfnamefont {H.}~\bibnamefont {Wang}}, \bibinfo {author} {\bibfnamefont {Y.-M.}\ \bibnamefont {He}}, \bibinfo {author} {\bibfnamefont {T.-H.}\ \bibnamefont {Chung}}, \bibinfo {author} {\bibfnamefont {H.}~\bibnamefont {Hu}}, \bibinfo {author} {\bibfnamefont {Y.}~\bibnamefont {Yu}}, \bibinfo {author} {\bibfnamefont {S.}~\bibnamefont {Chen}}, \bibinfo {author} {\bibfnamefont {X.}~\bibnamefont {Ding}}, \bibinfo {author} {\bibfnamefont {M.-C.}\ \bibnamefont {Chen}}, \bibinfo {author} {\bibfnamefont {J.}~\bibnamefont {Qin}}, \bibinfo {author} {\bibfnamefont {X.}~\bibnamefont {Yang}}, \emph {et~al.},\ }\bibfield  {title} {\bibinfo {title} {Towards optimal single-photon sources from polarized microcavities},\ }\href@noop {} {\bibfield  {journal} {\bibinfo  {journal} {Nature Photonics}\ }\textbf {\bibinfo {volume} {13}},\ \bibinfo {pages} {770} (\bibinfo {year} {2019})}\BibitemShut {NoStop}%
\bibitem [{\citenamefont {Brassard}\ \emph {et~al.}(2000)\citenamefont {Brassard}, \citenamefont {L{\"u}tkenhaus}, \citenamefont {Mor},\ and\ \citenamefont {Sanders}}]{brassard2000limitations}%
  \BibitemOpen
  \bibfield  {author} {\bibinfo {author} {\bibfnamefont {G.}~\bibnamefont {Brassard}}, \bibinfo {author} {\bibfnamefont {N.}~\bibnamefont {L{\"u}tkenhaus}}, \bibinfo {author} {\bibfnamefont {T.}~\bibnamefont {Mor}},\ and\ \bibinfo {author} {\bibfnamefont {B.~C.}\ \bibnamefont {Sanders}},\ }\bibfield  {title} {\bibinfo {title} {Limitations on practical quantum cryptography},\ }\href@noop {} {\bibfield  {journal} {\bibinfo  {journal} {Physical review letters}\ }\textbf {\bibinfo {volume} {85}},\ \bibinfo {pages} {1330} (\bibinfo {year} {2000})}\BibitemShut {NoStop}%
\bibitem [{\citenamefont {L{\"u}tkenhaus}(2000)}]{lutkenhaus2000security}%
  \BibitemOpen
  \bibfield  {author} {\bibinfo {author} {\bibfnamefont {N.}~\bibnamefont {L{\"u}tkenhaus}},\ }\bibfield  {title} {\bibinfo {title} {Security against individual attacks for realistic quantum key distribution},\ }\href@noop {} {\bibfield  {journal} {\bibinfo  {journal} {Physical Review A}\ }\textbf {\bibinfo {volume} {61}},\ \bibinfo {pages} {052304} (\bibinfo {year} {2000})}\BibitemShut {NoStop}%
\bibitem [{\citenamefont {Braginsky}\ \emph {et~al.}(1980)\citenamefont {Braginsky}, \citenamefont {Vorontsov},\ and\ \citenamefont {Thorne}}]{braginsky1980quantum}%
  \BibitemOpen
  \bibfield  {author} {\bibinfo {author} {\bibfnamefont {V.~B.}\ \bibnamefont {Braginsky}}, \bibinfo {author} {\bibfnamefont {Y.~I.}\ \bibnamefont {Vorontsov}},\ and\ \bibinfo {author} {\bibfnamefont {K.~S.}\ \bibnamefont {Thorne}},\ }\bibfield  {title} {\bibinfo {title} {Quantum nondemolition measurements},\ }\href@noop {} {\bibfield  {journal} {\bibinfo  {journal} {Science}\ }\textbf {\bibinfo {volume} {209}},\ \bibinfo {pages} {547} (\bibinfo {year} {1980})}\BibitemShut {NoStop}%
\bibitem [{\citenamefont {Stucki}\ \emph {et~al.}(2005)\citenamefont {Stucki}, \citenamefont {Brunner}, \citenamefont {Gisin}, \citenamefont {Scarani},\ and\ \citenamefont {Zbinden}}]{stucki2005fast}%
  \BibitemOpen
  \bibfield  {author} {\bibinfo {author} {\bibfnamefont {D.}~\bibnamefont {Stucki}}, \bibinfo {author} {\bibfnamefont {N.}~\bibnamefont {Brunner}}, \bibinfo {author} {\bibfnamefont {N.}~\bibnamefont {Gisin}}, \bibinfo {author} {\bibfnamefont {V.}~\bibnamefont {Scarani}},\ and\ \bibinfo {author} {\bibfnamefont {H.}~\bibnamefont {Zbinden}},\ }\bibfield  {title} {\bibinfo {title} {Fast and simple one-way quantum key distribution},\ }\href@noop {} {\bibfield  {journal} {\bibinfo  {journal} {Applied Physics Letters}\ }\textbf {\bibinfo {volume} {87}} (\bibinfo {year} {2005})}\BibitemShut {NoStop}%
\bibitem [{\citenamefont {Hwang}(2003)}]{hwang2003quantum}%
  \BibitemOpen
  \bibfield  {author} {\bibinfo {author} {\bibfnamefont {W.-Y.}\ \bibnamefont {Hwang}},\ }\bibfield  {title} {\bibinfo {title} {Quantum key distribution with high loss: toward global secure communication},\ }\href@noop {} {\bibfield  {journal} {\bibinfo  {journal} {Physical review letters}\ }\textbf {\bibinfo {volume} {91}},\ \bibinfo {pages} {057901} (\bibinfo {year} {2003})}\BibitemShut {NoStop}%
\bibitem [{\citenamefont {Lo}\ \emph {et~al.}(2005{\natexlab{a}})\citenamefont {Lo}, \citenamefont {Ma},\ and\ \citenamefont {Chen}}]{lo2005decoy}%
  \BibitemOpen
  \bibfield  {author} {\bibinfo {author} {\bibfnamefont {H.-K.}\ \bibnamefont {Lo}}, \bibinfo {author} {\bibfnamefont {X.}~\bibnamefont {Ma}},\ and\ \bibinfo {author} {\bibfnamefont {K.}~\bibnamefont {Chen}},\ }\bibfield  {title} {\bibinfo {title} {Decoy state quantum key distribution},\ }\href@noop {} {\bibfield  {journal} {\bibinfo  {journal} {Physical review letters}\ }\textbf {\bibinfo {volume} {94}},\ \bibinfo {pages} {230504} (\bibinfo {year} {2005}{\natexlab{a}})}\BibitemShut {NoStop}%
\bibitem [{\citenamefont {Rosenberg}\ \emph {et~al.}(2007)\citenamefont {Rosenberg}, \citenamefont {Harrington}, \citenamefont {Rice}, \citenamefont {Hiskett}, \citenamefont {Peterson}, \citenamefont {Hughes}, \citenamefont {Lita}, \citenamefont {Nam},\ and\ \citenamefont {Nordholt}}]{rosenberg2007long}%
  \BibitemOpen
  \bibfield  {author} {\bibinfo {author} {\bibfnamefont {D.}~\bibnamefont {Rosenberg}}, \bibinfo {author} {\bibfnamefont {J.~W.}\ \bibnamefont {Harrington}}, \bibinfo {author} {\bibfnamefont {P.~R.}\ \bibnamefont {Rice}}, \bibinfo {author} {\bibfnamefont {P.~A.}\ \bibnamefont {Hiskett}}, \bibinfo {author} {\bibfnamefont {C.~G.}\ \bibnamefont {Peterson}}, \bibinfo {author} {\bibfnamefont {R.~J.}\ \bibnamefont {Hughes}}, \bibinfo {author} {\bibfnamefont {A.~E.}\ \bibnamefont {Lita}}, \bibinfo {author} {\bibfnamefont {S.~W.}\ \bibnamefont {Nam}},\ and\ \bibinfo {author} {\bibfnamefont {J.~E.}\ \bibnamefont {Nordholt}},\ }\bibfield  {title} {\bibinfo {title} {Long-distance decoy-state quantum key distribution in optical fiber},\ }\href@noop {} {\bibfield  {journal} {\bibinfo  {journal} {Physical review letters}\ }\textbf {\bibinfo {volume} {98}},\ \bibinfo {pages} {010503} (\bibinfo {year} {2007})}\BibitemShut {NoStop}%
\bibitem [{\citenamefont {Schmitt-Manderbach}\ \emph {et~al.}(2007)\citenamefont {Schmitt-Manderbach}, \citenamefont {Weier}, \citenamefont {F{\"u}rst}, \citenamefont {Ursin}, \citenamefont {Tiefenbacher}, \citenamefont {Scheidl}, \citenamefont {Perdigues}, \citenamefont {Sodnik}, \citenamefont {Kurtsiefer}, \citenamefont {Rarity} \emph {et~al.}}]{schmitt2007experimental}%
  \BibitemOpen
  \bibfield  {author} {\bibinfo {author} {\bibfnamefont {T.}~\bibnamefont {Schmitt-Manderbach}}, \bibinfo {author} {\bibfnamefont {H.}~\bibnamefont {Weier}}, \bibinfo {author} {\bibfnamefont {M.}~\bibnamefont {F{\"u}rst}}, \bibinfo {author} {\bibfnamefont {R.}~\bibnamefont {Ursin}}, \bibinfo {author} {\bibfnamefont {F.}~\bibnamefont {Tiefenbacher}}, \bibinfo {author} {\bibfnamefont {T.}~\bibnamefont {Scheidl}}, \bibinfo {author} {\bibfnamefont {J.}~\bibnamefont {Perdigues}}, \bibinfo {author} {\bibfnamefont {Z.}~\bibnamefont {Sodnik}}, \bibinfo {author} {\bibfnamefont {C.}~\bibnamefont {Kurtsiefer}}, \bibinfo {author} {\bibfnamefont {J.~G.}\ \bibnamefont {Rarity}}, \emph {et~al.},\ }\bibfield  {title} {\bibinfo {title} {Experimental demonstration of free-space decoy-state quantum key distribution over 144 km},\ }\href@noop {} {\bibfield  {journal} {\bibinfo  {journal} {Physical Review Letters}\ }\textbf {\bibinfo {volume} {98}},\ \bibinfo {pages} {010504} (\bibinfo {year} {2007})}\BibitemShut {NoStop}%
\bibitem [{\citenamefont {Rusca}\ \emph {et~al.}(2018)\citenamefont {Rusca}, \citenamefont {Boaron}, \citenamefont {Gr{\"u}nenfelder}, \citenamefont {Martin},\ and\ \citenamefont {Zbinden}}]{rusca2018finite}%
  \BibitemOpen
  \bibfield  {author} {\bibinfo {author} {\bibfnamefont {D.}~\bibnamefont {Rusca}}, \bibinfo {author} {\bibfnamefont {A.}~\bibnamefont {Boaron}}, \bibinfo {author} {\bibfnamefont {F.}~\bibnamefont {Gr{\"u}nenfelder}}, \bibinfo {author} {\bibfnamefont {A.}~\bibnamefont {Martin}},\ and\ \bibinfo {author} {\bibfnamefont {H.}~\bibnamefont {Zbinden}},\ }\bibfield  {title} {\bibinfo {title} {Finite-key analysis for the 1-decoy state qkd protocol},\ }\href@noop {} {\bibfield  {journal} {\bibinfo  {journal} {Applied Physics Letters}\ }\textbf {\bibinfo {volume} {112}} (\bibinfo {year} {2018})}\BibitemShut {NoStop}%
\bibitem [{\citenamefont {Vogl}\ \emph {et~al.}(2017)\citenamefont {Vogl}, \citenamefont {Lu},\ and\ \citenamefont {Lam}}]{vogl2017room}%
  \BibitemOpen
  \bibfield  {author} {\bibinfo {author} {\bibfnamefont {T.}~\bibnamefont {Vogl}}, \bibinfo {author} {\bibfnamefont {Y.}~\bibnamefont {Lu}},\ and\ \bibinfo {author} {\bibfnamefont {P.~K.}\ \bibnamefont {Lam}},\ }\bibfield  {title} {\bibinfo {title} {Room temperature single photon source using fiber-integrated hexagonal boron nitride},\ }\href@noop {} {\bibfield  {journal} {\bibinfo  {journal} {Journal of Physics D: Applied Physics}\ }\textbf {\bibinfo {volume} {50}},\ \bibinfo {pages} {295101} (\bibinfo {year} {2017})}\BibitemShut {NoStop}%
\bibitem [{\citenamefont {Morrison}\ \emph {et~al.}(2021)\citenamefont {Morrison}, \citenamefont {Rambach}, \citenamefont {Koong}, \citenamefont {Graffitti}, \citenamefont {Thorburn}, \citenamefont {Kar}, \citenamefont {Ma}, \citenamefont {Park}, \citenamefont {Song}, \citenamefont {Stoltz} \emph {et~al.}}]{morrison2021bright}%
  \BibitemOpen
  \bibfield  {author} {\bibinfo {author} {\bibfnamefont {C.~L.}\ \bibnamefont {Morrison}}, \bibinfo {author} {\bibfnamefont {M.}~\bibnamefont {Rambach}}, \bibinfo {author} {\bibfnamefont {Z.~X.}\ \bibnamefont {Koong}}, \bibinfo {author} {\bibfnamefont {F.}~\bibnamefont {Graffitti}}, \bibinfo {author} {\bibfnamefont {F.}~\bibnamefont {Thorburn}}, \bibinfo {author} {\bibfnamefont {A.~K.}\ \bibnamefont {Kar}}, \bibinfo {author} {\bibfnamefont {Y.}~\bibnamefont {Ma}}, \bibinfo {author} {\bibfnamefont {S.-I.}\ \bibnamefont {Park}}, \bibinfo {author} {\bibfnamefont {J.~D.}\ \bibnamefont {Song}}, \bibinfo {author} {\bibfnamefont {N.~G.}\ \bibnamefont {Stoltz}}, \emph {et~al.},\ }\bibfield  {title} {\bibinfo {title} {A bright source of telecom single photons based on quantum frequency conversion},\ }\href@noop {} {\bibfield  {journal} {\bibinfo  {journal} {Applied Physics Letters}\ }\textbf {\bibinfo {volume} {118}} (\bibinfo {year} {2021})}\BibitemShut {NoStop}%
\bibitem [{\citenamefont {Thomas}\ \emph {et~al.}(2021)\citenamefont {Thomas}, \citenamefont {Billard}, \citenamefont {Coste}, \citenamefont {Wein}, \citenamefont {Ollivier}, \citenamefont {Krebs}, \citenamefont {Taza{\"\i}rt}, \citenamefont {Harouri}, \citenamefont {Lemaitre}, \citenamefont {Sagnes} \emph {et~al.}}]{thomas2021bright}%
  \BibitemOpen
  \bibfield  {author} {\bibinfo {author} {\bibfnamefont {S.}~\bibnamefont {Thomas}}, \bibinfo {author} {\bibfnamefont {M.}~\bibnamefont {Billard}}, \bibinfo {author} {\bibfnamefont {N.}~\bibnamefont {Coste}}, \bibinfo {author} {\bibfnamefont {S.}~\bibnamefont {Wein}}, \bibinfo {author} {\bibfnamefont {H.}~\bibnamefont {Ollivier}}, \bibinfo {author} {\bibfnamefont {O.}~\bibnamefont {Krebs}}, \bibinfo {author} {\bibfnamefont {L.}~\bibnamefont {Taza{\"\i}rt}}, \bibinfo {author} {\bibfnamefont {A.}~\bibnamefont {Harouri}}, \bibinfo {author} {\bibfnamefont {A.}~\bibnamefont {Lemaitre}}, \bibinfo {author} {\bibfnamefont {I.}~\bibnamefont {Sagnes}}, \emph {et~al.},\ }\bibfield  {title} {\bibinfo {title} {Bright polarized single-photon source based on a linear dipole},\ }\href@noop {} {\bibfield  {journal} {\bibinfo  {journal} {Physical review letters}\ }\textbf {\bibinfo {volume} {126}},\ \bibinfo {pages} {233601} (\bibinfo {year} {2021})}\BibitemShut {NoStop}%
\bibitem [{\citenamefont {Morrison}\ \emph {et~al.}(2023)\citenamefont {Morrison}, \citenamefont {Pousa}, \citenamefont {Graffitti}, \citenamefont {Koong}, \citenamefont {Barrow}, \citenamefont {Stoltz}, \citenamefont {Bouwmeester}, \citenamefont {Jeffers}, \citenamefont {Oi}, \citenamefont {Gerardot} \emph {et~al.}}]{morrison2023single}%
  \BibitemOpen
  \bibfield  {author} {\bibinfo {author} {\bibfnamefont {C.~L.}\ \bibnamefont {Morrison}}, \bibinfo {author} {\bibfnamefont {R.~G.}\ \bibnamefont {Pousa}}, \bibinfo {author} {\bibfnamefont {F.}~\bibnamefont {Graffitti}}, \bibinfo {author} {\bibfnamefont {Z.~X.}\ \bibnamefont {Koong}}, \bibinfo {author} {\bibfnamefont {P.}~\bibnamefont {Barrow}}, \bibinfo {author} {\bibfnamefont {N.~G.}\ \bibnamefont {Stoltz}}, \bibinfo {author} {\bibfnamefont {D.}~\bibnamefont {Bouwmeester}}, \bibinfo {author} {\bibfnamefont {J.}~\bibnamefont {Jeffers}}, \bibinfo {author} {\bibfnamefont {D.~K.}\ \bibnamefont {Oi}}, \bibinfo {author} {\bibfnamefont {B.~D.}\ \bibnamefont {Gerardot}}, \emph {et~al.},\ }\bibfield  {title} {\bibinfo {title} {Single-emitter quantum key distribution over 175 km of fibre with optimised finite key rates},\ }\href@noop {} {\bibfield  {journal} {\bibinfo  {journal} {Nature Communications}\ }\textbf {\bibinfo {volume} {14}},\ \bibinfo {pages} {3573} (\bibinfo {year} {2023})}\BibitemShut {NoStop}%
\bibitem [{\citenamefont {Lim}\ \emph {et~al.}(2014)\citenamefont {Lim}, \citenamefont {Curty}, \citenamefont {Walenta}, \citenamefont {Xu},\ and\ \citenamefont {Zbinden}}]{lim2014concise}%
  \BibitemOpen
  \bibfield  {author} {\bibinfo {author} {\bibfnamefont {C.~C.~W.}\ \bibnamefont {Lim}}, \bibinfo {author} {\bibfnamefont {M.}~\bibnamefont {Curty}}, \bibinfo {author} {\bibfnamefont {N.}~\bibnamefont {Walenta}}, \bibinfo {author} {\bibfnamefont {F.}~\bibnamefont {Xu}},\ and\ \bibinfo {author} {\bibfnamefont {H.}~\bibnamefont {Zbinden}},\ }\bibfield  {title} {\bibinfo {title} {Concise security bounds for practical decoy-state quantum key distribution},\ }\href@noop {} {\bibfield  {journal} {\bibinfo  {journal} {Physical Review A}\ }\textbf {\bibinfo {volume} {89}},\ \bibinfo {pages} {022307} (\bibinfo {year} {2014})}\BibitemShut {NoStop}%
\bibitem [{\citenamefont {Lo}\ \emph {et~al.}(2005{\natexlab{b}})\citenamefont {Lo}, \citenamefont {Chau},\ and\ \citenamefont {Ardehali}}]{lo2005efficient}%
  \BibitemOpen
  \bibfield  {author} {\bibinfo {author} {\bibfnamefont {H.-K.}\ \bibnamefont {Lo}}, \bibinfo {author} {\bibfnamefont {H.~F.}\ \bibnamefont {Chau}},\ and\ \bibinfo {author} {\bibfnamefont {M.}~\bibnamefont {Ardehali}},\ }\bibfield  {title} {\bibinfo {title} {Efficient quantum key distribution scheme and a proof of its unconditional security},\ }\href@noop {} {\bibfield  {journal} {\bibinfo  {journal} {Journal of Cryptology}\ }\textbf {\bibinfo {volume} {18}},\ \bibinfo {pages} {133} (\bibinfo {year} {2005}{\natexlab{b}})}\BibitemShut {NoStop}%
\bibitem [{\citenamefont {All{\'e}aume}\ \emph {et~al.}(2004)\citenamefont {All{\'e}aume}, \citenamefont {Treussart}, \citenamefont {Courty},\ and\ \citenamefont {Roch}}]{alleaume2004photon}%
  \BibitemOpen
  \bibfield  {author} {\bibinfo {author} {\bibfnamefont {R.}~\bibnamefont {All{\'e}aume}}, \bibinfo {author} {\bibfnamefont {F.}~\bibnamefont {Treussart}}, \bibinfo {author} {\bibfnamefont {J.-M.}\ \bibnamefont {Courty}},\ and\ \bibinfo {author} {\bibfnamefont {J.-F.}\ \bibnamefont {Roch}},\ }\bibfield  {title} {\bibinfo {title} {Photon statistics characterization of a single-photon source},\ }\href@noop {} {\bibfield  {journal} {\bibinfo  {journal} {New Journal of physics}\ }\textbf {\bibinfo {volume} {6}},\ \bibinfo {pages} {85} (\bibinfo {year} {2004})}\BibitemShut {NoStop}%
\bibitem [{\citenamefont {Waks}\ \emph {et~al.}(2002)\citenamefont {Waks}, \citenamefont {Santori},\ and\ \citenamefont {Yamamoto}}]{waks2002security}%
  \BibitemOpen
  \bibfield  {author} {\bibinfo {author} {\bibfnamefont {E.}~\bibnamefont {Waks}}, \bibinfo {author} {\bibfnamefont {C.}~\bibnamefont {Santori}},\ and\ \bibinfo {author} {\bibfnamefont {Y.}~\bibnamefont {Yamamoto}},\ }\bibfield  {title} {\bibinfo {title} {Security aspects of quantum key distribution with sub-poisson light},\ }\href@noop {} {\bibfield  {journal} {\bibinfo  {journal} {Physical Review A}\ }\textbf {\bibinfo {volume} {66}},\ \bibinfo {pages} {042315} (\bibinfo {year} {2002})}\BibitemShut {NoStop}%
\bibitem [{\citenamefont {Wei}\ \emph {et~al.}(2013)\citenamefont {Wei}, \citenamefont {Wang}, \citenamefont {Zhang}, \citenamefont {Gao}, \citenamefont {Ma},\ and\ \citenamefont {Ma}}]{wei2013decoy}%
  \BibitemOpen
  \bibfield  {author} {\bibinfo {author} {\bibfnamefont {Z.}~\bibnamefont {Wei}}, \bibinfo {author} {\bibfnamefont {W.}~\bibnamefont {Wang}}, \bibinfo {author} {\bibfnamefont {Z.}~\bibnamefont {Zhang}}, \bibinfo {author} {\bibfnamefont {M.}~\bibnamefont {Gao}}, \bibinfo {author} {\bibfnamefont {Z.}~\bibnamefont {Ma}},\ and\ \bibinfo {author} {\bibfnamefont {X.}~\bibnamefont {Ma}},\ }\bibfield  {title} {\bibinfo {title} {Decoy-state quantum key distribution with biased basis choice},\ }\href@noop {} {\bibfield  {journal} {\bibinfo  {journal} {Scientific reports}\ }\textbf {\bibinfo {volume} {3}},\ \bibinfo {pages} {2453} (\bibinfo {year} {2013})}\BibitemShut {NoStop}%
\bibitem [{\citenamefont {Yu}\ \emph {et~al.}(2016)\citenamefont {Yu}, \citenamefont {Zhou},\ and\ \citenamefont {Wang}}]{yu2016reexamination}%
  \BibitemOpen
  \bibfield  {author} {\bibinfo {author} {\bibfnamefont {Z.-W.}\ \bibnamefont {Yu}}, \bibinfo {author} {\bibfnamefont {Y.-H.}\ \bibnamefont {Zhou}},\ and\ \bibinfo {author} {\bibfnamefont {X.-B.}\ \bibnamefont {Wang}},\ }\bibfield  {title} {\bibinfo {title} {Reexamination of decoy-state quantum key distribution with biased bases},\ }\href@noop {} {\bibfield  {journal} {\bibinfo  {journal} {Physical Review A}\ }\textbf {\bibinfo {volume} {93}},\ \bibinfo {pages} {032307} (\bibinfo {year} {2016})}\BibitemShut {NoStop}%
\bibitem [{\citenamefont {Devetak}\ and\ \citenamefont {Winter}(2005)}]{devetak2005distillation}%
  \BibitemOpen
  \bibfield  {author} {\bibinfo {author} {\bibfnamefont {I.}~\bibnamefont {Devetak}}\ and\ \bibinfo {author} {\bibfnamefont {A.}~\bibnamefont {Winter}},\ }\bibfield  {title} {\bibinfo {title} {Distillation of secret key and entanglement from quantum states},\ }\href@noop {} {\bibfield  {journal} {\bibinfo  {journal} {Proceedings of the Royal Society A: Mathematical, Physical and engineering sciences}\ }\textbf {\bibinfo {volume} {461}},\ \bibinfo {pages} {207} (\bibinfo {year} {2005})}\BibitemShut {NoStop}%
\bibitem [{\citenamefont {Hasegawa}\ \emph {et~al.}(2007)\citenamefont {Hasegawa}, \citenamefont {Hayashi}, \citenamefont {Hiroshima},\ and\ \citenamefont {Tomita}}]{hasegawa2007security}%
  \BibitemOpen
  \bibfield  {author} {\bibinfo {author} {\bibfnamefont {J.}~\bibnamefont {Hasegawa}}, \bibinfo {author} {\bibfnamefont {M.}~\bibnamefont {Hayashi}}, \bibinfo {author} {\bibfnamefont {T.}~\bibnamefont {Hiroshima}},\ and\ \bibinfo {author} {\bibfnamefont {A.}~\bibnamefont {Tomita}},\ }\bibfield  {title} {\bibinfo {title} {Security analysis of decoy state quantum key distribution incorporating finite statistics},\ }\href@noop {} {\bibfield  {journal} {\bibinfo  {journal} {arXiv preprint arXiv:0707.3541}\ } (\bibinfo {year} {2007})}\BibitemShut {NoStop}%
\bibitem [{\citenamefont {Ma}\ \emph {et~al.}(2005)\citenamefont {Ma}, \citenamefont {Qi}, \citenamefont {Zhao},\ and\ \citenamefont {Lo}}]{ma2005practical}%
  \BibitemOpen
  \bibfield  {author} {\bibinfo {author} {\bibfnamefont {X.}~\bibnamefont {Ma}}, \bibinfo {author} {\bibfnamefont {B.}~\bibnamefont {Qi}}, \bibinfo {author} {\bibfnamefont {Y.}~\bibnamefont {Zhao}},\ and\ \bibinfo {author} {\bibfnamefont {H.-K.}\ \bibnamefont {Lo}},\ }\bibfield  {title} {\bibinfo {title} {Practical decoy state for quantum key distribution},\ }\href@noop {} {\bibfield  {journal} {\bibinfo  {journal} {Physical Review A}\ }\textbf {\bibinfo {volume} {72}},\ \bibinfo {pages} {012326} (\bibinfo {year} {2005})}\BibitemShut {NoStop}%
\bibitem [{\citenamefont {Curty}\ \emph {et~al.}(2014)\citenamefont {Curty}, \citenamefont {Xu}, \citenamefont {Cui}, \citenamefont {Lim}, \citenamefont {Tamaki},\ and\ \citenamefont {Lo}}]{curty2014finite}%
  \BibitemOpen
  \bibfield  {author} {\bibinfo {author} {\bibfnamefont {M.}~\bibnamefont {Curty}}, \bibinfo {author} {\bibfnamefont {F.}~\bibnamefont {Xu}}, \bibinfo {author} {\bibfnamefont {W.}~\bibnamefont {Cui}}, \bibinfo {author} {\bibfnamefont {C.~C.~W.}\ \bibnamefont {Lim}}, \bibinfo {author} {\bibfnamefont {K.}~\bibnamefont {Tamaki}},\ and\ \bibinfo {author} {\bibfnamefont {H.-K.}\ \bibnamefont {Lo}},\ }\bibfield  {title} {\bibinfo {title} {Finite-key analysis for measurement-device-independent quantum key distribution},\ }\href@noop {} {\bibfield  {journal} {\bibinfo  {journal} {Nature communications}\ }\textbf {\bibinfo {volume} {5}},\ \bibinfo {pages} {3732} (\bibinfo {year} {2014})}\BibitemShut {NoStop}%
\bibitem [{\citenamefont {Yin}\ \emph {et~al.}(2020)\citenamefont {Yin}, \citenamefont {Zhou}, \citenamefont {Gu}, \citenamefont {Xie}, \citenamefont {Lu},\ and\ \citenamefont {Chen}}]{yin2020tight}%
  \BibitemOpen
  \bibfield  {author} {\bibinfo {author} {\bibfnamefont {H.-L.}\ \bibnamefont {Yin}}, \bibinfo {author} {\bibfnamefont {M.-G.}\ \bibnamefont {Zhou}}, \bibinfo {author} {\bibfnamefont {J.}~\bibnamefont {Gu}}, \bibinfo {author} {\bibfnamefont {Y.-M.}\ \bibnamefont {Xie}}, \bibinfo {author} {\bibfnamefont {Y.-S.}\ \bibnamefont {Lu}},\ and\ \bibinfo {author} {\bibfnamefont {Z.-B.}\ \bibnamefont {Chen}},\ }\bibfield  {title} {\bibinfo {title} {Tight security bounds for decoy-state quantum key distribution},\ }\href@noop {} {\bibfield  {journal} {\bibinfo  {journal} {Scientific Reports}\ }\textbf {\bibinfo {volume} {10}},\ \bibinfo {pages} {1} (\bibinfo {year} {2020})}\BibitemShut {NoStop}%
\bibitem [{\citenamefont {Sidhu}\ \emph {et~al.}(2022)\citenamefont {Sidhu}, \citenamefont {Brougham}, \citenamefont {McArthur}, \citenamefont {Pousa},\ and\ \citenamefont {Oi}}]{sidhu2022finite}%
  \BibitemOpen
  \bibfield  {author} {\bibinfo {author} {\bibfnamefont {J.~S.}\ \bibnamefont {Sidhu}}, \bibinfo {author} {\bibfnamefont {T.}~\bibnamefont {Brougham}}, \bibinfo {author} {\bibfnamefont {D.}~\bibnamefont {McArthur}}, \bibinfo {author} {\bibfnamefont {R.~G.}\ \bibnamefont {Pousa}},\ and\ \bibinfo {author} {\bibfnamefont {D.~K.}\ \bibnamefont {Oi}},\ }\bibfield  {title} {\bibinfo {title} {Finite key effects in satellite quantum key distribution},\ }\href@noop {} {\bibfield  {journal} {\bibinfo  {journal} {npj Quantum Information}\ }\textbf {\bibinfo {volume} {8}},\ \bibinfo {pages} {18} (\bibinfo {year} {2022})}\BibitemShut {NoStop}%
\bibitem [{\citenamefont {Cai}\ and\ \citenamefont {Scarani}(2009)}]{cai2009finite}%
  \BibitemOpen
  \bibfield  {author} {\bibinfo {author} {\bibfnamefont {R.~Y.}\ \bibnamefont {Cai}}\ and\ \bibinfo {author} {\bibfnamefont {V.}~\bibnamefont {Scarani}},\ }\bibfield  {title} {\bibinfo {title} {Finite-key analysis for practical implementations of quantum key distribution},\ }\href@noop {} {\bibfield  {journal} {\bibinfo  {journal} {New Journal of Physics}\ }\textbf {\bibinfo {volume} {11}},\ \bibinfo {pages} {045024} (\bibinfo {year} {2009})}\BibitemShut {NoStop}%
\bibitem [{\citenamefont {Shor}\ and\ \citenamefont {Preskill}(2000)}]{shor2000simple}%
  \BibitemOpen
  \bibfield  {author} {\bibinfo {author} {\bibfnamefont {P.~W.}\ \bibnamefont {Shor}}\ and\ \bibinfo {author} {\bibfnamefont {J.}~\bibnamefont {Preskill}},\ }\bibfield  {title} {\bibinfo {title} {Simple proof of security of the bb84 quantum key distribution protocol},\ }\href@noop {} {\bibfield  {journal} {\bibinfo  {journal} {Physical review letters}\ }\textbf {\bibinfo {volume} {85}},\ \bibinfo {pages} {441} (\bibinfo {year} {2000})}\BibitemShut {NoStop}%
\bibitem [{\citenamefont {Tomamichel}\ \emph {et~al.}(2017)\citenamefont {Tomamichel}, \citenamefont {Martinez-Mateo}, \citenamefont {Pacher},\ and\ \citenamefont {Elkouss}}]{tomamichel2017fundamental}%
  \BibitemOpen
  \bibfield  {author} {\bibinfo {author} {\bibfnamefont {M.}~\bibnamefont {Tomamichel}}, \bibinfo {author} {\bibfnamefont {J.}~\bibnamefont {Martinez-Mateo}}, \bibinfo {author} {\bibfnamefont {C.}~\bibnamefont {Pacher}},\ and\ \bibinfo {author} {\bibfnamefont {D.}~\bibnamefont {Elkouss}},\ }\bibfield  {title} {\bibinfo {title} {Fundamental finite key limits for one-way information reconciliation in quantum key distribution},\ }\href@noop {} {\bibfield  {journal} {\bibinfo  {journal} {Quantum Information Processing}\ }\textbf {\bibinfo {volume} {16}},\ \bibinfo {pages} {1} (\bibinfo {year} {2017})}\BibitemShut {NoStop}%
\bibitem [{\citenamefont {Renner}(2008)}]{renner2008security}%
  \BibitemOpen
  \bibfield  {author} {\bibinfo {author} {\bibfnamefont {R.}~\bibnamefont {Renner}},\ }\bibfield  {title} {\bibinfo {title} {Security of quantum key distribution},\ }\href@noop {} {\bibfield  {journal} {\bibinfo  {journal} {International Journal of Quantum Information}\ }\textbf {\bibinfo {volume} {6}},\ \bibinfo {pages} {1} (\bibinfo {year} {2008})}\BibitemShut {NoStop}%
\bibitem [{\citenamefont {Dixon}\ \emph {et~al.}(2008)\citenamefont {Dixon}, \citenamefont {Yuan}, \citenamefont {Dynes}, \citenamefont {Sharpe},\ and\ \citenamefont {Shields}}]{dixon2008gigahertz}%
  \BibitemOpen
  \bibfield  {author} {\bibinfo {author} {\bibfnamefont {A.}~\bibnamefont {Dixon}}, \bibinfo {author} {\bibfnamefont {Z.}~\bibnamefont {Yuan}}, \bibinfo {author} {\bibfnamefont {J.}~\bibnamefont {Dynes}}, \bibinfo {author} {\bibfnamefont {A.}~\bibnamefont {Sharpe}},\ and\ \bibinfo {author} {\bibfnamefont {A.}~\bibnamefont {Shields}},\ }\bibfield  {title} {\bibinfo {title} {Gigahertz decoy quantum key distribution with 1 mbit/s secure key rate},\ }\href@noop {} {\bibfield  {journal} {\bibinfo  {journal} {Optics express}\ }\textbf {\bibinfo {volume} {16}},\ \bibinfo {pages} {18790} (\bibinfo {year} {2008})}\BibitemShut {NoStop}%
\bibitem [{\citenamefont {Rakhlin}\ \emph {et~al.}(2023)\citenamefont {Rakhlin}, \citenamefont {Galimov}, \citenamefont {Dyakonov}, \citenamefont {Skryabin}, \citenamefont {Klimko}, \citenamefont {Kulagina}, \citenamefont {Zadiranov}, \citenamefont {Sorokin}, \citenamefont {Sedova}, \citenamefont {Guseva} \emph {et~al.}}]{rakhlin2023demultiplexed}%
  \BibitemOpen
  \bibfield  {author} {\bibinfo {author} {\bibfnamefont {M.}~\bibnamefont {Rakhlin}}, \bibinfo {author} {\bibfnamefont {A.}~\bibnamefont {Galimov}}, \bibinfo {author} {\bibfnamefont {I.}~\bibnamefont {Dyakonov}}, \bibinfo {author} {\bibfnamefont {N.}~\bibnamefont {Skryabin}}, \bibinfo {author} {\bibfnamefont {G.}~\bibnamefont {Klimko}}, \bibinfo {author} {\bibfnamefont {M.}~\bibnamefont {Kulagina}}, \bibinfo {author} {\bibfnamefont {Y.~M.}\ \bibnamefont {Zadiranov}}, \bibinfo {author} {\bibfnamefont {S.}~\bibnamefont {Sorokin}}, \bibinfo {author} {\bibfnamefont {I.}~\bibnamefont {Sedova}}, \bibinfo {author} {\bibfnamefont {Y.~A.}\ \bibnamefont {Guseva}}, \emph {et~al.},\ }\bibfield  {title} {\bibinfo {title} {Demultiplexed single-photon source with a quantum dot coupled to microresonator},\ }\href@noop {} {\bibfield  {journal} {\bibinfo  {journal} {Journal of Luminescence}\ }\textbf {\bibinfo {volume} {253}},\ \bibinfo {pages} {119496} (\bibinfo {year} {2023})}\BibitemShut {NoStop}%
\bibitem [{\citenamefont {Wang}\ \emph {et~al.}(2009)\citenamefont {Wang}, \citenamefont {Yang}, \citenamefont {Peng},\ and\ \citenamefont {Pan}}]{wang2009decoy}%
  \BibitemOpen
  \bibfield  {author} {\bibinfo {author} {\bibfnamefont {X.-B.}\ \bibnamefont {Wang}}, \bibinfo {author} {\bibfnamefont {L.}~\bibnamefont {Yang}}, \bibinfo {author} {\bibfnamefont {C.-Z.}\ \bibnamefont {Peng}},\ and\ \bibinfo {author} {\bibfnamefont {J.-W.}\ \bibnamefont {Pan}},\ }\bibfield  {title} {\bibinfo {title} {Decoy-state quantum key distribution with both source errors and statistical fluctuations},\ }\href@noop {} {\bibfield  {journal} {\bibinfo  {journal} {New Journal of Physics}\ }\textbf {\bibinfo {volume} {11}},\ \bibinfo {pages} {075006} (\bibinfo {year} {2009})}\BibitemShut {NoStop}%
\end{thebibliography}%

\end{document}